\documentclass[aps,amsfonts,amsmath,prd,nofootinbib,tightenlines]{revtex4}

\usepackage{hyperref}
\usepackage{epsfig}
\usepackage{bm}
\usepackage{bbm}
\newcommand{\wavenumbersq}{k^2}
\newcommand{\wavenumber}{k}
\newcommand{\wavevars}{\wavenumber}
\renewcommand{\Im}{{\mathop{\rm Im}}}
\renewcommand{\Re}{{\mathop{\rm Re}}}

\begin{document}
\title{Vacuum Polarization Energy of a Nonzero Radius Cosmic String
Induced Spacetime}


\author{N. Graham$^{a)}$, H. Weigel$^{b)}$}

\affiliation{
$^{a)}$Department of Physics, Middlebury College
Middlebury, VT 05753, USA\\
$^{b)}$Institute for Theoretical Physics, Physics Department,
Stellenbosch University, Matieland 7602, South Africa}

\begin{abstract}
We calculate the vacuum polarization energy (VPE) of a scalar field in
the background of a nonsingular cosmic string in $2+1$ spacetime
dimensions.  Our calculation expresses the VPE as a renormalized sum
and integral over scattering data, which can be expressed in terms of
Legendre and Bessel functions for the ``ballpoint pen'' model,
analogous to a square well in curvature, and which can be obtained
numerically for a generic string profile.  We show how relationships
between the local density of states, expressed in terms of the Green's
function, and the global density of states, expressed in terms of the Jost
function, extend to this curved spacetime background and allow for
precise implementation of perturbative renormalization conditions.
\end{abstract}

\maketitle

\section{Introduction}

While there does not exist a fully quantized theory of gravity,
it does appear possible to extend quantum field theory to a classical
curved spacetime background \cite{BrlDv}, although subtleties such as
Hawking radiation and its consequences for black hole information
remain unresolved.  Examples in which one can compute the
contributions of quantum fluctuations to local densities are therefore
especially valuable for gaining insight into the interplay between
quantum mechanics and general relativity.  In the case of the
Schwarzschild metric, calculations using scattering Green's functions
such as \cite{Candelas,Candelas2,PhysRevLett.70.1739,PhysRevD.51.4337}
make it possible to demonstrate the emergence of Hawking radiation via
a physically motivated choice of vacuum.  However, such geometries
inevitably contain singular points, making it difficult to consider
global quantities that would be obtained via integration over space of
local densities, and one is often restricted to calculations of local
densities in regions with no matter.

In this paper we consider a simpler geometry in two space dimensions,
consisting of a localized defect leading to a deficit angle at large
distances.  We consider a defect with a core of nonzero radius,
leading to a nonsingular configuration in which we can consider
regions both with and without matter.  In higher dimensions, this
configuration would extend to a cosmic string geometry.  We focus
especially on the ``ballpoint pen'' model
\cite{PhysRevD.31.3288,1985ApJ...288..422G,PhysRevD.42.2669}, in which
a region of constant curvature is smoothly joined to a flat region
with a deficit angle, forming the analog of a square well in
curvature, and compute the total vacuum polarization energy (VPE) of a
scalar field in this background.  Local densities in this model were
computed previously in \cite{PhysRevD.110.105009,PhysRevD.111.105028}
which in principle could be integrated over space to obtain the
results found here.  However, this integral converges only as a
principal value at the step boundary, so it is not tractable to
compute it numerically.  Instead, we show how to use an analytic
relationship between the integral over space of the Green's function
and the Jost function from potential scattering theory to circumvent
the need to compute this integral, leading to a fully renormalized
calculation in which divergences are removed via local counterterms
fixed by standard renormalization conditions.  Our calculation draws
on scattering data derived in Refs.\
\cite{PhysRevD.59.064017,Khusnutdinov:2004ux}, but our results differ
because we use only these standard subtractions to implement renormalization.

We begin by introducing the background geometry and scattering problem
in Sec.\ \ref{sec:model}.  We then show in Sec.\ \ref{sec:dos} how to
connect the local density of states, defined via the Green's function,
to the global density of states, defined via the Jost function, which
is then used to compute the VPE.  We carry out this calculation for
the ballpoint pen model in Sec.\ \ref{sec:vpe}, and discuss our results
in Sec.\ \ref{sec:results}.  In Appendix \ref{sec:variable} we explain
how to extend all of these results to an arbitrary string profile
using a variable phase calculation and in Appendix \ref{sec:analytic}
we discuss aspects of analytic continuation in wave number $k$ and
angular momentum $\ell$.

\section{Model}

\label{sec:model}

We consider a free scalar field $\phi$ of mass $\mu$ in $2+1$ spacetime
dimensions, for which the action functional is
\begin{equation}
S = -\frac{1}{2}\int d^3 x \sqrt{-g} \left(
\nabla_\alpha \phi \nabla^\alpha \phi + \xi {\cal R} \phi^2
+ \mu^2 \phi^2\right) \,,
\end{equation}
including a coupling to the Ricci curvature scalar ${\cal R}$ with
strength  $\xi$.  Of
particular interest is the case of conformal coupling, $\displaystyle
\xi = \frac{1}{8}$ in two space dimensions.  The equation of motion is
\begin{equation}
-\nabla_\alpha\nabla^\alpha\phi + \xi {\cal R} \phi + \mu^2 \phi= 0
\end{equation}
and we consider the spacetime metric
\cite{PhysRevD.31.3288,1985ApJ...288..422G,PhysRevD.42.2669}
\begin{equation}
ds^2 = -dt^2 + p(r)^2 dr^2 + r^2 d\theta^2
\end{equation}
with a deficit angle $2\theta_0$, meaning that the
range of angular coordinate is $\theta = 0\ldots 2(\pi-\theta_0)$, and we
define $\displaystyle \sigma = \frac{\pi}{\pi-\theta_0}$.  To
implement the deficit angle without a singularity at the origin, we
introduce a profile function $p(r)$ that ranges from 
$\displaystyle \frac{1}{\sigma}$ at the origin
to $1$ at the string radius $r_0$ and beyond.  The
nonzero Christoffel symbols in this geometry are
\cite{PhysRevD.31.3288,1985ApJ...288..422G,PhysRevD.42.2669}
\begin{equation}
\Gamma_{r r}^r =  \frac{p'(r)}{p(r)}\,, \qquad
\Gamma_{\theta \theta}^r = -\frac{r}{p(r)^2}\,, 
\qquad \hbox{and} \qquad
\Gamma_{\theta r}^\theta = \Gamma_{r \theta}^\theta = \frac{1}{r}\,,
\end{equation}
and because the geometry only has curvature in two dimensions, all the
nonzero components of the Riemann and Ricci tensors
\begin{equation}
R_{\theta r\theta}^r = -R_{\theta\theta r}^r = 
R_{\theta\theta} = g_{\theta\theta} \frac{{\cal R}}{2}
\qquad \hbox{and} \qquad
R_{r\theta r}^\theta = -R_{rr\theta}^\theta
=  R_{rr} = g_{rr} \frac{{\cal R}}{2}
\label{eqn:curvature}
\end{equation}
can be expressed in terms of the curvature scalar $\displaystyle
{\cal R}=\frac{2}{r} \frac{p'(r)}{p(r)^3}$.  It obeys the
Gauss-Bonnet theorem in two dimensions,
\begin{equation}
\int_0^{r_0} p(r) dr \int_0^{2\pi/\sigma} r d\theta \, {\cal R}
= \frac{4\pi}{\sigma} \int_0^{r_0} \frac{p'(r)}{p(r)^2} dr =
\left.-\frac{4\pi}{\sigma} \frac{1}{p(r)} \right|_{r=0}^{r=r_0}
= 4\pi \left(1-\frac{1}{\sigma}\right) = 4 \theta_0 \,,
\end{equation}
for any $p(r)$ with the boundary conditions given above.

Acting on any scalar $\chi$, the covariant derivatives simply become
ordinary derivatives, while for second derivatives we have nontrivial
contributions from the Christoffel symbols,
\begin{equation}
\nabla_\theta \nabla_\theta \chi = \partial_\theta^2 \chi -
\Gamma_{\theta\theta}^r \partial_r \chi \,, \qquad
\nabla_r \nabla_r \chi = \partial_r^2 \chi -
\Gamma_{rr}^r \partial_r \chi \,,
\qquad \hbox{and} \qquad
\nabla_r \nabla_\theta \chi =
\nabla_\theta \nabla_r \chi = \partial_\theta \partial_r \chi -
\Gamma_{r\theta}^\theta \partial_\theta \chi \,,
\end{equation}
and so covariant derivatives with respect to $\theta$
can be nonzero even if $\chi$ is rotationally invariant.  In
particular, we have the Laplace-Beltrami operator
\begin{equation}
(g^{\theta \theta} \nabla_\theta \nabla_\theta + g^{rr} \nabla_r
\nabla_r) \chi = \frac{1}{r^2} \left(\frac{\partial^2 \chi}
{\partial \theta^2} + {\cal D}_r^2 \right)\chi,
\end{equation}
where $\displaystyle {\cal D}_r = \frac{r}{p(r)} \frac{\partial}{\partial r}$
is the radial derivative.  The equation of motion for $\phi$ then becomes
\begin{equation}
\left(\frac{\partial^2}{\partial t^2} - \frac{1}{r^2} {\cal D}_r^2
 - \frac{1}{r^2} \frac{\partial^2}{\partial \theta^2} 
+ \mu^2 + \xi {\cal R}\right) \phi = 0\,.
\label{eqn:eomphi}
\end{equation}

We define the Green's function $G(\bm{r},\bm{r}',\wavevars)$
for wave number $k$, which obeys
\begin{equation}
\left(-\frac{1}{r^2} {\cal D}_r^2
- \frac{1}{r^2} \frac{\partial^2}{\partial \theta^2}
+ \xi {\cal R} - \wavenumbersq \right)
G(\bm{r},\bm{r}',\wavevars)  = \frac{1}{r p(r)} 
\delta(r-r') \delta(\theta-\theta')\,,
\end{equation}
and we furthermore define the changes in radius and area as
\begin{equation}
\Delta r = \int_0^\infty (p(r)-1)\,dr
\end{equation}
and
\begin{equation}
\Delta A = \frac{2\pi}{\sigma} \int_0^\infty r (p(r)-1)\,dr \,,
\label{eqn:area}
\end{equation}
respectively.

We can then decompose the Green's function in a partial wave basis as
\begin{equation}
G(\bm{r},\bm{r}',\wavevars) = 
\sum_{\ell=0}^\infty{}' G_\ell(\bm{r},\bm{r}',\wavevars) = 
\frac{i\sigma}{2} \sum_{\ell=0}^\infty{}'
\psi_{\wavevars,\ell}^{\rm reg} (r_<) \psi_{\wavevars,\ell}^{\rm out} (r_>)
\cos \left[\sigma \ell (\theta-\theta')\right] \,,
\label{eqn:green}
\end{equation}
where the subscripts indicate the larger and smaller of the two radii and
the prime indicates that the $\ell=0$ term in the sum is counted
with a weight of one-half, which arises because we restrict to $\ell
\geq 0$.  This sum is written in terms of the regular and
outgoing scattering wavefunctions, which are solutions to the radial
wave equation
\begin{equation}
\left(-\frac{1}{r^2} {\cal D}_r^2 + 
\frac{\ell^2\sigma^2}{r^2} + \xi {\cal R}
\right)\psi_{\ell,k}(r) = k^2 \psi_{\ell,k}(r) \,,
\label{eqn:eom}
\end{equation}
normalized according to the Wronskian relation
\begin{equation}
\frac{d}{dr} \left(\psi_{\wavevars,\ell}^{\rm reg}(r)\right)
\psi_{\wavevars,\ell}^{\rm out}(r)
- \psi_{\wavevars,\ell}^{\rm reg}(r) \frac{d}{dr} 
\left(\psi_{\wavevars,\ell}^{\rm out}(r)\right)
= \frac{p(r)}{r} \,,
\label{eqn:wronk}
\end{equation}
where the outgoing wavefunction approaches the Hankel function
$H^{(1)}_{\sigma \ell} (kr)$ at large distances.

We consider the ``ballpoint pen'' profile function 
\cite{PhysRevD.31.3288,1985ApJ...288..422G,PhysRevD.42.2669}
\begin{equation}
p^{\rm pen}(r) = 
\begin{cases}
\left[\sigma^2-\frac{r^2}{r_0^2} (\sigma^2-1)\right]^{-1/2}&r<r_0\\
1 &r>r_0\end{cases}\,,
\end{equation}
where $r_0$ is the string radius, which gives constant curvature
$\displaystyle {\cal R}=\frac{2 (\sigma^2-1)}{r_0^2}$
inside and zero curvature outside.  It has geometrical factors
\begin{equation}
\Delta r^{\rm pen} = \left(\frac{\arctan \sqrt{\sigma^2-1}}
{\sqrt{\sigma^2-1}} - 1\right) r_0
\qquad \hbox{and} \qquad
\Delta A^{\rm pen} = -\frac{\pi r_0^2}{\sigma} \frac{\sigma-1}{\sigma+1}
\end{equation}
and its scattering wavefunctions are given in terms of the Legendre
functions of the first and second kind
$P_{\nu(\wavevars)}^{-\ell}$ and $Q_{\nu(\wavevars)}^{\ell}$ 
and the Bessel and Hankel functions $J_{\sigma \ell}$ and 
$H^{(1)}_{\sigma \ell}$ by
\begin{equation}
\psi_{\wavevars,\ell}^{\rm pen}(r) =
\begin{array}{|c|c|c|}
\hline
& r<r_0 & r>r_0\cr
\hline
\hbox{regular} & A_{\wavevars,\ell}^{\rm pen}
P_{\nu(\wavevars)}^{-\ell}\left(\tfrac{1}{\sigma  p^{\rm pen}(r)}\right) & 
J_{\sigma \ell}(\wavenumber r) + B_{\wavevars,\ell}^{\rm pen}
H^{(1)}_{\sigma \ell}(\wavenumber r)
\cr
\hline
\hbox{outgoing} &  
C_{\wavevars,\ell}^{\rm pen}
P_{\nu(\wavevars)}^{-\ell}\left(\tfrac{1}{\sigma p^{\rm pen}(r)}\right) +
D_{\wavevars,\ell}^{\rm pen} 
Q_{\nu(\wavevars)}^{\ell}\left(\tfrac{1}{\sigma p^{\rm pen}(r)}\right)
& H^{(1)}_{\sigma \ell}(\wavenumber r) \cr
\hline
\end{array}
\end{equation}
with
\begin{equation}
\nu(\wavevars) = -\frac{1}{2} + \frac{1}{2}\sqrt{(1-8\xi) + 
\frac{4 \wavenumbersq r_0^2}{\sigma^2-1}} \,.
\end{equation}
and
\begin{eqnarray}
A_{\wavevars,\ell}^{\rm pen} D_{\wavevars,\ell}^{\rm pen} &=& 
\frac{2i(-1)^\ell}{\pi\sigma} \\
\frac{C_{\wavevars,\ell}^{\rm pen}}{D_{\wavevars,\ell}^{\rm pen}} &=& -
\frac{
(\sigma^2-1)Q_{\nu(\wavevars)}^{\ell}{}'\left(\tfrac{1}{\sigma}\right)
H^{(1)}_{\sigma \ell}\left(\wavenumber r_0\right) +
\sigma \wavenumber r_0 Q_{\nu(\wavevars)}^{\ell}\left(\tfrac{1}{\sigma}\right)
H^{(1)}_{\sigma \ell}{}'\left(\wavenumber r_0\right)
}{
(\sigma^2-1)P_{\nu(\wavevars)}^{-\ell}{}'\left(\tfrac{1}{\sigma}\right)
H^{(1)}_{\sigma \ell}\left(\wavenumber r_0\right) +
\sigma \wavenumber r_0 P_{\nu(\wavevars)}^{-\ell}\left(\tfrac{1}{\sigma}\right)
H^{(1)}_{\sigma \ell}{}'\left(\wavenumber r_0\right)
}
\cr
B_{\wavevars,\ell}^{\rm pen} &=& -\frac{
(\sigma^2-1)P_{\nu(\wavevars)}^{-\ell}{}'\left(\tfrac{1}{\sigma}\right)
J_{\sigma \ell}\left(\wavenumber r_0\right) +
\sigma \wavenumber r_0 P_{\nu(\wavevars)}^{-\ell}\left(\tfrac{1}{\sigma}\right)
J_{\sigma \ell}'\left(\wavenumber r_0\right)
}{
(\sigma^2-1)P_{\nu(\wavevars)}^{-\ell}{}'\left(\tfrac{1}{\sigma}\right)
H^{(1)}_{\sigma \ell}\left(\wavenumber r_0\right) +
\sigma \wavenumber r_0 P_{\nu(\wavevars)}^{-\ell}\left(\tfrac{1}{\sigma}\right)
H^{(1)}_{\sigma \ell}{}'\left(\wavenumber r_0\right)
}
\end{eqnarray}
where a prime denotes a derivative with respect to the function's
argument and we have assumed $\ell \geq 0$.  While all four
coefficients can be obtained by matching the wavefunction and its
first derivative at $r=r_0$, here we show only the combinations
of these coefficients that we will need;  the product
$A_{\wavevars,\ell}^{\rm pen} D_{\wavevars,\ell}^{\rm pen}$ is
particularly simple due to the Wronskian relation.  We can then
extract the Jost function $F_\ell(k)$
\cite{Chadan:1977pq,Newton:1982qc} from the 
$B_{\wavevars,\ell}^{\rm pen}$  coefficient as \cite{PhysRevD.59.064017}
\begin{equation}
F^{\rm pen}_\ell(k) = 
\frac{(k r_0)^\ell}{(\sigma^2-1)^{\ell/2}}
\frac{i^{\ell(\sigma-1)}\pi}{2 i \sigma^{3/2}}\left[
(\sigma^2-1)P_{\nu(\wavevars)}^{-\ell}{}'\left(\tfrac{1}{\sigma}\right)
H^{(1)}_{\sigma \ell}\left(\wavenumber r_0\right) +
\sigma \wavenumber r_0 P_{\nu(\wavevars)}^{-\ell}\left(\tfrac{1}{\sigma}\right)
H^{(1)}_{\sigma \ell}{}'\left(\wavenumber r_0\right) \right]
\end{equation}
by using
\begin{equation}
1 + 2 B_{\wavevars,\ell}^{\rm pen} = \frac{F_\ell(-k)}{F_\ell(k)} \,,
\end{equation}
where the normalization is chosen to give the asymptotic limit
\begin{equation}
\lim_{k\to\infty} F_\ell(k) e^{ik \Delta r} = 1
\end{equation}
with $F(-k) = F(k)^*$ for $k$ real.  We define the $S$-matrix
\begin{equation}
S_\ell(k) = \frac{F(-k)}{F(k)}\,,
\end{equation}
which obeys
\begin{equation}
\lim_{k \to \infty} e^{-2ik \Delta r} S_\ell(k) = 1 \hbox{\quad and \quad}
\lim_{\ell \to \infty} (-1)^{\ell(\sigma-1)} S_\ell(k) = 1 \,.
\end{equation}
We note that it is also straightforward to extract analogous
scattering data for the ``flowerpot'' model, in which the curvature is
concentrated at a $\delta$-function ring of radius $r_0$.  However,
such a configuration has divergent total energy due to this sharp limit
\cite{density}.  In Appendix \ref{sec:variable} we explain how to obtain
scattering data numerically for an arbitrary continuous profile $p(r)$.

\section{Point String and Density of States}

\label{sec:dos}

In order to pass from local densities, which can be computed from the
Green's function, to global quantities like the total energy, we
connect the integral over space of the Green's function to the Jost
function.  More specifically, $\displaystyle \frac{2k}{i}$ times
the imaginary part of the Green's function gives a local density of
states, and the corresponding global density of states is given 
by $\displaystyle \frac{1}{\pi}$ times the $k$-derivative of the Jost
function \cite{density}.  Since the global density of states diverges
with volume, we must consider differences, for example by subtracting
the corresponding results for empty space, to obtain a meaningful
result.  In curved space, this subtraction is complicated by the
changes to the background geometry.  

To form a simple comparison, we consider the ``point string''
configuration
\cite{PhysRevD.34.1918,PhysRevD.35.536,PhysRevD.35.3779}, which is the
$r_0\to 0$ limit of the ballpoint pen.  It consists of
flat spacetime with the same deficit angle for $r>0$ and the
associated curvature concentrated at $r=0$.  We can similarly
construct its Green's function from the wavefunctions
\begin{equation}
\psi_{\wavevars,\ell}^{\rm point}(r) =
\begin{array}{|c|c|}
\hline
\hbox{regular} & J_{\sigma \ell}\left(\wavenumber r\right) \cr
\hline
\hbox{outgoing} &  H^{(1)}_{\sigma \ell}(\wavenumber r) \cr
\hline
\end{array}
\end{equation}
and we obtain, for $\ell \geq 0$,
\begin{equation}
\frac{2k}{i \pi} \int_0^\infty \frac{2 \pi r}{\sigma} dr \left(
G_\ell(r,r,k) p(r) - G_\ell^{\rm point}(r,r,k)\right)
= \frac{i}{\pi} \frac{d}{dk} \log \left((i k r_0)^{\ell(\sigma-1)}
F_\ell(k)\right)
\label{eqn:dos}
\end{equation}
for complex $k$ in the upper half-plane, where the nonzero imaginary
part of $k$ is necessary for convergence of the integral on the
left-hand side.  This formula connects the
local and global density of states, generalizing to curved spacetime
the relationship between the spatial integral of the Green's function
and the Jost function derived in Ref.\ \cite{density}.  In the limit
where $k$ approaches the real axis from above, the real part of this
equation represents the density of states in channel $\ell$,
\begin{equation}
\Delta \rho_\ell(k) = 
\rho_\ell(k) - \rho^{\rm point}_\ell(k) = \frac{1}{\pi} \frac{d}{dk}
\delta_\ell(k) 
\end{equation}
where we have defined the phase shift as
\begin{equation}
\delta_\ell(k) = \frac{1}{2i} \log S_\ell
= - \arg F_\ell(k)
\label{eqn:phase}
\end{equation}
for $k$ real.

On dimensional grounds, the vacuum energy density of the point
string goes like $\displaystyle \frac{1}{r^3}$ for $\mu=0$, giving a
divergent total energy.  Nonetheless, we will be able to use it as an
intermediate step when renormalizing the VPE of the ballpoint pen or
other nonsingular configurations.

\section{Vacuum Polarization Energy}

\label{sec:vpe}

The VPE can be constructed as a sum over the
zero-point energies $\displaystyle \frac{1}{2} \sqrt{k^2 + \mu^2}$ for
all the modes of oscillation of the scalar field.  To put this sum in a
tractable form for calculation in an unbounded region, we rewrite it
as an integral weighted by the continuum density of states, as
calculated above.  (We would also need to include the explicit
contribution from bound states, but there are none for the
configurations we consider.)  Such calculations are typically divergent,
requiring that we introduce local counterterms with coefficients
defined through perturbative renormalization conditions.  First, we
must make a free space (cosmological constant) subtraction to account
for the change in area, which multiplies the change in the free density of
states in two dimensions $\displaystyle \Delta \rho^{\rm free}(k) = 2
\sum_{\ell=0}^\infty {}' \Delta \rho_\ell^{\rm free}(k) =
\frac{k\Delta A}{2\pi}$ for $k\geq 0$.  This
subtraction represents the difference in total energy
between the point string and empty space, which is divergent when
integrated over $k$.  As we will see below, when we combine this
quantity with the difference between the nonzero radius string and the
point string, which has a corresponding divergence as described in the
previous section, we obtain a finite result.

The only other potential divergence in two space dimensions is the
tadpole counterterm, the coefficient of which is given by the integral
over space of $\displaystyle \frac{\cal R}{48\pi}$
\cite{PhysRevD.110.105009,PhysRevD.111.105028}.  Its
contribution is given by replacing the phase shift defined in Eq.\
(\ref{eqn:phase}) by its first Born approximation \cite{density}.
However, when summed over $\ell$, this quantity is independent of $k$,
since it is dimensionless and must be proportional to the
dimensionless quantity  $\displaystyle \int_0^\infty r p(r) 
{\cal R}(r) \, dr$, with no other dependence on the string background,
and it is independent of $\mu$. As a result, a unique feature of two
dimensions is that the tadpole term gives no contribution to the
energy after taking the $k$ derivative.

Putting these results together, we can compute the full VPE on the
real $k$ axis,
\begin{eqnarray}
{\cal E} &=& \int_0^\infty \frac{\sqrt{k^2+\mu^2}}{2\pi}  \left\{
2\sum_{\ell=0}^\infty {}' \left[\frac{1}{2i} \frac{d}{dk} \log\left(
(-1)^{\ell(\sigma-1)} S_\ell(k)
\right)\right]  - \frac{k}{2}\Delta A \right\} dk \cr
&=& \Re \int_0^\infty \frac{\sqrt{k^2+\mu^2}}{2\pi}  \left\{
2\sum_{\ell=0}^\infty {}' \left[
i \frac{d}{dk} \log \left((i k r_0)^{\ell(\sigma-1)} F_\ell(k)\right)
- \frac{k \Delta A}{2\pi} \frac{1}{i(\ell+\frac{1}{2})}\right] 
- \frac{k}{2}\Delta A \right\} dk\,,
\label{eqn:vpe}
\end{eqnarray}
where the subtraction of the second term in brackets in the second
line does not affect the real part but renders the integral
convergent.  In the latter form, the integrand is analytic in the
upper half-plane, allowing us to draw on the density of states
relation in Eq.\ (\ref{eqn:dos}) to ensure that it correctly describes
the VPE.

\section{Results and Conclusions}

\label{sec:results}

\begin{figure}[htbp]
\includegraphics[width=0.5\linewidth]{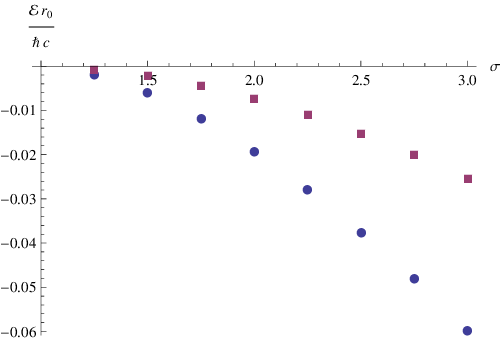}
\caption{Total vacuum polarization energy (VPE) ${\cal E}$ in units of
$\displaystyle \frac{1}{r_0}$ of the ballpoint pen string for a scalar
field with mass $\mu=0$, as a function of $\sigma$, computed using
Eq.\ (\ref{eqn:vpe}).  The sum over $\ell$ and integral over $k$ have
been truncated and then extrapolated to infinity.
Circles are for minimal coupling ($\xi = 0$)
while squares are for conformal coupling ($\displaystyle \xi =
\frac{1}{8}$).}
\label{fig:energy}
\end{figure}

By substituting the Jost function and geometrical factors for the
ballpoint pen and choosing mass $\mu=0$, using Eq.\ (\ref{eqn:vpe})
we obtain the results for the total energy shown in
Figure~\ref{fig:energy}.  Since $r_0$ is the
only scale in the problem for a massless field, the energy must be
proportional to $\displaystyle \frac{1}{r_0}$, so that $r_0 {\cal E}$
is a function of $\sigma$ and $\xi$ only.  Results for both minimal
coupling ($\xi = 0$) and conformal coupling ($\displaystyle \xi =
\frac{1}{8}$) are shown as functions of $\sigma$.  We see that in both
cases the total energy is everywhere negative and roughly proportional
to $-\sigma(\sigma^2-1)$.  As expected from Ref.\
\cite{PhysRevD.110.105009}, the energy is smaller in magnitude for
conformal coupling than minimal coupling, but not zero in either case.

We find results of a similar order of magnitude compared to those in
Refs.\ \cite{PhysRevD.59.064017,Khusnutdinov:2004ux} for the limit
where $\sigma$ is close to $1$, but our results differ quantitatively
because we subtract only local counterterms rather than the more
complex quantities extracted from uniform asymptotic expansions and
zeta function regularization in those calculations.  In particular,
our results appear to vanish as $\sigma^2-1$ rather than
$(\sigma^2-1)^2$ as $\sigma$ approaches $1$.

The integrand in the second line of Eq.\ (\ref{eqn:vpe}) is an
analytic function of $k$ in the upper half-plane, a property we
used to make the connection in Eq.\ (\ref{eqn:dos}) between the local
density of states as given by the Green's function and the global
density of states as given by the Jost function.  As a result, we
might hope to use contour integration to rewrite the VPE as an
integral on the imaginary axis $k=i\kappa$ arising from the cut in the
$\sqrt{k^2 + \mu^2}$ term, as is standard in flat-space calculations
\cite{Bordag:1994jz,density,Graham:2022rqk}.  However, we find that
this integral does not converge.  Perhaps related, to first extend the
integral to the entire real axis before closing the contour, one
typically takes advantage of the derivative of the phase shift being
even, but here there are also odd terms arising from the point string
and the free density of states, suggesting that $k$ should be replaced
by $\sqrt{k^2}$ in these terms.

If it is possible to extend the calculation to the imaginary axis, an
additional benefit would be to make it possible to use Regge techniques
to replace the sum over $\ell$ by an integral over imaginary $\ell = i
\lambda$, which can avoid problems associated with the mismatch
between sums over $\ell$ and $\sigma \ell$ in renormalization
\cite{PhysRevD.110.105009,PhysRevD.111.105028}.  As explained in
Appendix\ \ref{sec:analytic}, this replacement can be carried out 
successfully in our calculation, but, as with other examples of this
technique, only after analytic continuation in $k$.

\acknowledgments
N.\ G.\ is supported in part by the
National Science Foundation (NSF) through grant PHY-2209582.
H.\ W.\ is supported in part by the National Research Foundation of
South Africa (NRF) by grant~150672.

\appendix

\section{Variable Phase Method}
\label{sec:variable}

If we instead consider an arbitrary smooth profile function $p(r)$ that
interpolates between 
$\displaystyle \frac{1}{\sigma}$ at the origin and $1$ at infinity, we
can develop general expressions for the Green's function, Jost
function, and $S$-matrix using the variable phase method.

We define the physical distance $r_*$, which is the solution to the
ordinary differential equation $\displaystyle
\frac{dr_*}{dr} = p(r)$ with $r_*=r=0$ at the origin.  We can then
rewrite the wave equation in terms of the reduced wavefunction
$\phi(r_*) = \sqrt{r} \psi(r)$ as
\cite{PhysRevD.59.064017,PhysRevD.110.105009}
\begin{equation}
\left[-\frac{d^2}{dr_*^2}  + 
\frac{\sigma^2 \ell^2-\frac{1}{4 p(r)^2}}{r^2} + 
\left(\xi-\frac{1}{4}\right) {\cal R} \right]\phi_{\ell,k}(r_*) 
= k^2 \phi_{\ell,k}(r_*) \,,
\end{equation}
where $r$ is now a function of $r_*$.

We define the potential including the centrifugal angular momentum
term as
\begin{equation}
V_\ell(r_*) = 
\frac{\sigma^2 \ell^2-\frac{1}{4 p(r)^2}}{r^2} 
+ \left(\xi-\frac{1}{4}\right) {\cal R} \,,
\end{equation}
and note that near the origin, $\displaystyle p(r) \approx
\frac{1}{\sigma}$ and $r \approx  \sigma r_*$,  and so the
centrifugal term corresponds to angular momentum $\ell$, while at
large distances $p(r) \to 1$ and $r \to r_* + \Delta r$
so the centrifugal term has angular momentum $\sigma \ell$.  

We define the regular and outgoing variable phase
solutions  \cite{density} to Eq.\ (\ref{eqn:eom}) in terms of a Hankel
function with a Riccati-type rescaling $\hat H^{(1)}_\ell (z) =
\sqrt{z} H^{(1)}_\ell (z)$ by writing
\begin{eqnarray}
\phi^{\text{\small reg}}_{\ell,k}(r_*) &=& \frac{h_{\ell,k}(r_*)}
{\hat H^{(1)}_\ell (k r_*)} \cr
\phi^{\rm out}_{\ell,k}(r_*) &=& g_{\ell,k}(r_*) 
\hat H^{(1)}_{\sigma \ell} (k r_*)\,,
\end{eqnarray}
which obey the differential equations
\begin{eqnarray}
-h''_{\ell,k}(r_*) 
+2 k \frac{d}{dr_*} \left(
\frac{\hat H^{(1)}_\ell{}' (k r_*)}{\hat H^{(1)}_\ell (k r_*)} h_{\ell,k}(r_*) \right) 
+ \left(V_\ell(r_*) - \frac{\ell^2 - \frac{1}{4}}{r_*^2}\right)
h_{\ell,k}(r_*) = 0 \cr
-g''_{\ell,k}(r_*) - 2 k \frac{\hat H^{(1)}_{\sigma \ell}{}' (k r_*)}
{\hat H^{(1)}_{\sigma \ell} (k r_*)} g'_{\ell,k}(r_*)
+ \left(V_\ell(r_*) - \frac{\sigma^2 \ell^2 - \frac{1}{4}}{r_*^2}\right) g_{\ell,k}(r_*)
= 0 \, ,
\end{eqnarray}
with boundary conditions $\displaystyle \lim_{r_* \to 0}
h_{\ell,k}(r_*) = \displaystyle \lim_{r_* \to \infty} g_{\ell,k}'(r_*) = 0$,
$\displaystyle \lim_{r_* \to 0} h_{\ell,k}'(r_*) = 1$, and $\displaystyle
\lim_{r_* \to \infty} g_{\ell,k}(r_*) = i^{(\sigma-1)\ell}$,
where a prime denotes derivative with respect to the argument
and the factor of $i^{(\sigma-1)\ell}$ accounts for the
difference in phase between $\hat H^{(1)}_{\sigma\ell}(k r_*)$ and
$\hat H^{(1)}_\ell(k r_*)$.  In practice, we must start the
integration for $h_{\ell,k}(r_*)$ with these boundary conditions imposed
at some small value just larger than zero.

We can then compute the Wronskian of these two solutions,
\begin{eqnarray}
{\cal W}_\ell(k) &=& 
\frac{\hat H^{(1)}_{\sigma \ell} (k r_*)}{\hat H^{(1)}_\ell (k r_*)} \Bigg[
\left(h'_{\ell,k}(r_*) - k \frac{\hat H^{(1)}_\ell{}' (k r_*)}{\hat
H^{(1)}_\ell (k r_*)} h_{\ell,k}(r_*) \right) g_{\ell,k}(r_*) \cr
&& 
- h_{\ell,k}(r_*)
\left(g'_{\ell,k}(r_*) + k \frac{\hat H^{(1)}_{\sigma \ell}{}' (k r_*)}
{\hat H^{(1)}_{\sigma\ell} (k r_*)} g_{\ell,k}(r_*) \right) \Bigg]\,,
\label{eqn:wronkgen}
\end{eqnarray}
which is independent of $r_*$, so it can be evaluated at an
intermediate fitting point of our choice.  For fixed $\ell$, the Wronskian
goes to $1$ as $k \to \infty$ and obeys 
${\cal W}_\ell(-k) = {\cal W}_\ell(k)^*$ for $k$ real.  The Jost
function is then given by $F_\ell(k) = e^{-ik \Delta r}
{\cal W}_\ell(k)$, and from these results we obtain the Green's
function as
\begin{equation}
G(\bm{r},\bm{r}',k)  = \frac{\sigma}{\pi} \frac{1}{\sqrt{r r'}} 
\sum_{\ell=0}^\infty{}' \frac{\hat H^{(1)}_{\sigma\ell} (k {r_*}_>)}
{\hat H^{(1)}_{\ell} (k {r_*}_<)} \frac{h_{\ell,k}({r_*}_<) 
g_{\ell,k}({r_*}_>)}{{\cal W}_\ell(k)} \cos [\sigma\ell(\theta-\theta')] \,.
\label{eqn:greengen}
\end{equation}
These expressions can then be used to calculate
the total energy as in the main text, as well as to compute local
densities  as described in Refs.\
\cite{PhysRevD.110.105009,PhysRevD.111.105028}.  This approach also extends
directly to complex $k$ in the upper half-plane \cite{density}.

\section{Complex Angular Momentum}
\label{sec:analytic}

The deficit angle background complicates standard techniques for
computing quantum corrections because it leads to a quantized angular
momentum $\sigma \ell$ for integer $\ell$, which must be compared to
angular momentum $\ell$ for the free background.  As a result, one
often does not obtain term-by-term cancellation in the resulting sum
over $\ell$.  In some cases, such as conical geometries
\cite{Maghrebi6867,physics5040065} and the point string
\cite{PhysRevD.110.105009}, this problem can be addressed by
replacing the discrete $\ell$ sum with a continuous
Kontorovich-Lebedev integral over imaginary $\ell$, where the wave
number $k$ is also analytically continued to imaginary values
\cite{Oberhettinger,KL}.

Our analytic scattering data results for the ballpoint pen extend
directly to complex $\ell$, as does the density of states relation in
Eq.\ (\ref{eqn:dos}).  For quantities involving sums of analytic
functions over $\ell$, in particular Eq.\ (\ref{eqn:green}) or the
second line of Eq.\ (\ref{eqn:vpe}), breaking the cosine into complex
exponentials and then multiplying by $\displaystyle \frac{\pi \exp(\pm i \pi
\ell)}{\sin \pi \ell}$, which has poles of unit residue at integer $\ell$,
allows us to equate the original sum to a contour integral in complex
$\ell$ over a semicircle in the right half-plane.  In the limit of
large radius, the contribution of the semicircular arc goes to zero,
leaving only the contribution on the imaginary axis, which by symmetry
can then be simplified to an integral over the positive imaginary axis only.
The Green's function in Eq.\ (\ref{eqn:green}) thus becomes
\cite{PhysRevD.110.105009}
\begin{equation}
G(\bm{r},\bm{r}',i\kappa) = \frac{1}{2\pi} \int_0^\infty 
\frac{id\lambda}{\sinh \frac{\lambda \pi}{\sigma}}
\cosh \left[\lambda\left(\frac{\pi}{\sigma} -
|\theta-\theta'|\right)\right]
\left[\psi_{i\kappa,i\frac{\lambda}{\sigma}}^{\rm reg} (r_<)
\psi_{i\kappa,i\frac{\lambda}{\sigma}}^{\rm out} (r_>) -
\psi_{i\kappa,-i\frac{\lambda}{\sigma}}^{\rm reg} (r_<) 
\psi_{i\kappa,-i\frac{\lambda}{\sigma}}^{\rm out} (r_>) \right] \,,
\label{eqn:greencontin}
\end{equation}
with $i\lambda = \ell \sigma$ and $k=i\kappa$.
We note that the difference of regular functions becomes
proportional to the outgoing function, so that the jump condition arises
via the angular dependence while the radial dependence is continuous.
Similarly, we find for the analytic continuation of the integrand in
Eq.\ (\ref{eqn:vpe}),
\begin{equation}
\sum_{\ell=0}^\infty {}' \left[
\frac{d}{d\kappa} \log \left((\kappa r_0)^{\ell(\sigma-1)} 
F_\ell(i\kappa)\right)
- \frac{\kappa \Delta A}{2\pi\left(\ell+\frac{1}{2}\right)}\right] 
= \Im \int_0^\infty \frac{\coth \frac{\pi\lambda}{\sigma} d\lambda}{\sigma}  
\left[
\frac{\kappa \Delta A}{2\pi\left(\frac{i\lambda}{\sigma}+\frac{1}{2}\right)} -
\frac{d}{d\kappa} \log \left((\kappa r_0)^{\frac{i\lambda(\sigma-1)}{\sigma}} 
F_{\frac{i\lambda}{\sigma}}(i\kappa)\right)
\right],
\label{eqn:integrand}
\end{equation}
where again $i\lambda = \ell \sigma$ and $k=i\kappa$.  We have checked
that these integrals agree with the original sums, but because
they are applicable only for imaginary $k=i\kappa$, further
modification to close the contour integral would be needed to apply
them to the full calculation of the energy. 

We also note that when $\ell$ is not a real integer, for $r<r_0$ in
the ballpoint pen model it is computationally preferable to take as
independent solutions
$P_{\nu(\wavevars)}^{-\ell}$ and $P_{\nu(\wavevars)}^{\ell}$  
rather than 
$P_{\nu(\wavevars)}^{-\ell}$ and $Q_{\nu(\wavevars)}^{\ell}$.
With this replacement made throughout, the expressions for the
scattering data hold as above, except that the product of the
$A_{\wavevars,\ell}$ and $D_{\wavevars,\ell}$ coefficients becomes
$\displaystyle \frac{\pi} {2 \sigma \sin \pi \ell}$.

We can also extend the variable phase method of Appendix
\ref{sec:variable} to complex $\ell$.  The calculation given
there is valid when $\ell$ is an integer (though it also gives the
correct result for $\ell$ any positive real number).  To carry out the
analytic continuation in $\ell$, we must use a modified approach that
is instead valid only when $\ell$ is not an integer.  In this case,
the calculation of the regular wave depends much more sensitively on
the boundary conditions at the origin.  As a result, we must replace
the simpler boundary conditions on $h_{\ell,k}(r_*)$ given in Appendix
\ref{sec:variable} by those obtained from the expansion of the free
regular solution
\begin{equation}
h_{\ell,k}(r_*) \approx 
r_*\left[1 + \left(\frac{kr_*}{2i}\right)^{2 \ell}\frac{\Gamma(-\ell)}
{\Gamma(\ell)} \right]
\end{equation}
and we must furthermore introduce overall factors of $\displaystyle
\frac{1}{2\ell}$ in the expressions for the Wronskian in Eq.\
(\ref{eqn:wronkgen}) and the Green's function in Eq.\
(\ref{eqn:greengen}), which then cancel in the latter equation.  
(After introducing these same factors of $\displaystyle
\frac{1}{2\ell}$, we can similarly use the boundary condition
$h_{\ell,k}(r_*) \approx r_*$ for the case of nonzero integer $\ell$
discussed in Appendix \ref{sec:variable}, along with the
limit described in Eq.\ (\ref{eqn:ellzero}) below for the case of
$\ell=0$.)  We thus obtain the Green's function as
\begin{eqnarray}
G(\bm{r},\bm{r}',i\kappa) &=& \frac{\sigma}{2\pi^2\sqrt{r r'}} \int_0^\infty 
\frac{d\lambda}{i \lambda \sinh \frac{\lambda \pi}{\sigma}}
\cosh \left[\lambda\left(\frac{\pi}{\sigma} -
|\theta-\theta'|\right)\right] 
\cr && \hspace*{1.25cm} \times
\left[\frac{
h_{\frac{i\lambda}{\sigma},i\kappa}({r_*}_<)
g_{\frac{i\lambda}{\sigma},i\kappa}({r_*}_>)
\hat H^{(1)}_{i\lambda} (i\kappa {r_*}_>)}
{\hat H^{(1)}_{\frac{i\lambda}{\sigma}} (i\kappa {r_*}_<)
{\cal W}_\frac{i\lambda}{\sigma}(i\kappa)}
-  \frac{h_{-\frac{i\lambda}{\sigma},i\kappa}({r_*}_<) 
g_{-\frac{i\lambda}{\sigma},i\kappa}({r_*}_>)
\hat H^{(1)}_{-i\lambda} (i\kappa {r_*}_>)}
{\hat H^{(1)}_{-\frac{i\lambda}{\sigma}}(i\kappa {r_*}_<)
{\cal W}_{-\frac{i\lambda}{\sigma}}(i\kappa)} \right],
\label{eqn:greencontin2}
\end{eqnarray}
in the same way as in Eq.\ (\ref{eqn:greencontin}), where the
Green's function is now defined as the right-hand side of Eq.\
(\ref{eqn:greengen}) divided by $2\ell$
and the Wronskian is now defined as the right-hand side of Eq.\
(\ref{eqn:wronkgen}) divided by $2\ell$, giving results that obey
Eqs.\ (\ref{eqn:dos}) and (\ref{eqn:integrand}).  While it appears
that these rescalings could be avoided by appropriate redefinitions,
they become necessary when $\Re ~ \ell = 0$.  Such modifications are
analogous to the procedure in Ref.\ \cite{density} for the $\ell
= 0$ case, which uses the limit
\begin{equation}
\lim_{\ell\to0}\frac{r_*}{2\ell}
\left[1+\left(\frac{kr_*}{2i}\right)^{2\ell}\frac{\Gamma(-\ell)}{\Gamma(\ell)}\right]
=-r_*\ln\left(\frac{kr_*}{2i}\right)+\mathcal{O}(r_*)
\label{eqn:ellzero}
\end{equation}
to obtain the behavior of the regular solution $h_{\ell=0,k}(r_*)$ at
the origin.

\bibliographystyle{apsrev}
\bibliography{flowerpen}

@ARTICLE{Candelas, AUTHOR  =  "P. Candelas",
                 TITLE   =  "Vacuum polarization in {Schwarzschild} spacetime",
                 JOURNAL =  "Phys. Rev. D",
                 YEAR    =  "1980",
                 VOLUME  =  "21",
                 NUMBER  =  "8",
                 PAGES   =  "2185",
                 MONTH   =  "April",
                 NOTE    =  ""}

@article{Candelas2,
  title = {Vacuum $<{\ensuremath{\varphi}}^{2}>$ in Schwarzschild spacetime},
  author = {Candelas, P. and Howard, K. W.},
  journal = {Phys. Rev. D},
  volume = {29},
  issue = {8},
  pages = {1618--1625},
  numpages = {0},
  year = {1984},
  month = {Apr},
  publisher = {American Physical Society},
  doi = {10.1103/PhysRevD.29.1618}
}

@article{density,
      author         = "Graham, N. and Jaffe, R.L. and Khemani, V. and Quandt, M.
                        and Scandurra, M. and Weigel, H.",
      title          = "{Calculating vacuum energies in renormalizable quantum
                        field theories: A New approach to the Casimir problem}",
      journal        = "Nucl. Phys. B",
      volume         = "645",
      pages          = "49-84",
      doi            = "10.1016/S0550-3213(02)00823-4",
      year           = "2002",
      //eprint         = "hep-th/0207120",
      archivePrefix  = "arXiv",
      primaryClass   = "hep-th",
}

@BOOK{BrlDv, AUTHOR    =  "N. D. Birrell and P. C. W. Davies",
              TITLE     =  "Quantum Fields in Curved Space",
              PUBLISHER =  "Cambridge University Press",
              YEAR      =  "1982",
              VOLUME    =  "",
              SERIES    =  "Cambridge Monographs on Mathematical Physics",
              ADDRESS   =  "Cambridge",
              EDITION   =  "",
              MONTH     =  "",
              NOTE      =  ""}

@article{PhysRevD.110.105009,
  title = {Quantum energy density of cosmic strings with nonzero radius},
  author = {Koike, Mao and Laquidain, Xabier and Graham, Noah},
  journal = {Phys. Rev. D},
  volume = {110},
  issue = {10},
  pages = {105009},
  numpages = {11},
  year = {2024},
  month = {Nov},
  publisher = {American Physical Society},
  doi = {10.1103/PhysRevD.110.105009},
  //url = {https://link.aps.org/doi/10.1103/PhysRevD.110.105009}
}

@article{PhysRevD.111.105028,
  title = {Renormalized quantum stress-energy tensor of a nonzero radius cosmic string},
  author = {Graham, Noah},
  journal = {Phys. Rev. D},
  volume = {111},
  issue = {10},
  pages = {105028},
  numpages = {12},
  year = {2025},
  month = {May},
  publisher = {American Physical Society},
  doi = {10.1103/8zvv-prwk},
  //url = {https://link.aps.org/doi/10.1103/8zvv-prwk}
}

@article{Oberhettinger,
author = {Oberhettinger, F.},
title = {Diffraction of waves by a wedge},
journal = {Communications on Pure and Applied Mathematics},
volume = {7},
number = {3},
pages = {551-563},
doi = {https://doi.org/10.1002/cpa.3160070306},
//url = {https://onlinelibrary.wiley.com/doi/abs/10.1002/cpa.3160070306},
//eprint = {https://onlinelibrary.wiley.com/doi/pdf/10.1002/cpa.3160070306},
year = {1954}
}

@article{PhysRevD.59.064017,
  title = {Ground state energy of a massive scalar field in the background of a cosmic string of finite thickness},
  author = {Khusnutdinov, N. R. and Bordag, M.},
  journal = {Phys. Rev. D},
  volume = {59},
  issue = {6},
  pages = {064017},
  numpages = {12},
  year = {1999},
  month = {Feb},
  publisher = {American Physical Society},
  //url = {https://link.aps.org/doi/10.1103/PhysRevD.59.064017},
  doi = {10.1103/PhysRevD.59.064017}
}

@article{Khusnutdinov:2004ux,
    author = "Khusnutdinov, N. R. and Khabibullin, A. R.",
    title = "{Zero point energy of a massless scalar field in the cosmic string space-time}",
    doi = "10.1023/B:GERG.0000032153.48990.3f",
    journal = "Gen. Rel. Grav.",
    volume = "36",
    pages = "1613--1618",
    year = "2004"
}

@article{PhysRevD.42.2669,
  title = {Effects of curvature couplings for quantum fields on cosmic-string space-times},
  author = {Allen, Bruce and Ottewill, Adrian C.},
  journal = {Phys. Rev. D},
  volume = {42},
  issue = {8},
  pages = {2669--2677},
  numpages = {0},
  year = {1990},
  month = {Oct},
  publisher = {American Physical Society},
  //url = {https://link.aps.org/doi/10.1103/PhysRevD.42.2669},
  doi = {10.1103/PhysRevD.42.2669}
}

@article{PhysRevD.35.536,
  title = {Quantum field theory in the space-time of a cosmic string},
  author = {Linet, B.},
  journal = {Phys. Rev. D},
  volume = {35},
  issue = {2},
  pages = {536--539},
  numpages = {0},
  year = {1987},
  month = {Jan},
  publisher = {American Physical Society},
  //url = {https://link.aps.org/doi/10.1103/PhysRevD.35.536},
  doi = {10.1103/PhysRevD.35.536}
}

@article{PhysRevD.35.3779,
  title = {Vacuum polarization in the gravitational field of a cosmic string},
  author = {Frolov, V. P. and Serebriany, E. M.},
  journal = {Phys. Rev. D},
  volume = {35},
  issue = {12},
  pages = {3779--3782},
  numpages = {0},
  year = {1987},
  month = {Jun},
  publisher = {American Physical Society},
  //url = {https://link.aps.org/doi/10.1103/PhysRevD.35.3779},
  doi = {10.1103/PhysRevD.35.3779}
}

@article{PhysRevD.34.1918,
  title = {Vacuum fluctuations outside cosmic strings},
  author = {Helliwell, T. M. and Konkowski, D. A.},
  journal = {Phys. Rev. D},
  volume = {34},
  issue = {6},
  pages = {1918--1920},
  numpages = {0},
  year = {1986},
  month = {Sep},
  publisher = {American Physical Society},
  //url = {https://link.aps.org/doi/10.1103/PhysRevD.34.1918},
  doi = {10.1103/PhysRevD.34.1918}
}

@Article{physics5040065,
AUTHOR = {Graham, Noah},
TITLE = {Electromagnetic Casimir–Polder Interaction for a Conducting Cone},
JOURNAL = {Physics},
VOLUME = {5},
YEAR = {2023},
NUMBER = {4},
PAGES = {1003--1012},
//URL = {https://www.mdpi.com/2624-8174/5/4/65},
//ISSN = {2624-8174},
ABSTRACT = {Using the formulation of the electromagnetic Green's function of a perfectly conducting cone in terms of analytically continued angular momentum, we compute the Casimir–Polder interaction energy of a cone with a polarizable particle. We introduce this formalism by first reviewing the analogous approach for a perfectly conducting wedge, and then demonstrate the calculation through numerical evaluation of the resulting integrals.},
DOI = {10.3390/physics5040065}
}

@article{Maghrebi6867,
	Author = {Maghrebi, Mohammad F. and Rahi, Sahand Jamal and Emig, Thorsten and Graham, Noah and Jaffe, Robert L. and Kardar, Mehran},
	Doi = {10.1073/pnas.1018079108},
	//Eprint = {https://www.pnas.org/content/108/17/6867.full.pdf},
	//Issn = {0027-8424},
	Journal = {Proceedings of the National Academy of Sciences},
	Number = {17},
	Pages = {6867--6871},
	Publisher = {National Academy of Sciences},
	Title = {Analytical results on Casimir forces for conductors with edges and tips},
	//Url = {https://www.pnas.org/content/108/17/6867},
	Volume = {108},
	Year = {2011}}

@article{PhysRevD.31.3288,
  title = {Exact gravitational field of a string},
  author = {Hiscock, William A.},
  journal = {Phys. Rev. D},
  volume = {31},
  issue = {12},
  pages = {3288--3290},
  numpages = {0},
  year = {1985},
  month = {Jun},
  publisher = {American Physical Society},
  //url = {https://link.aps.org/doi/10.1103/PhysRevD.31.3288},
  doi = {10.1103/PhysRevD.31.3288}
}

@ARTICLE{1985ApJ...288..422G,
       author = {{Gott III}, J.~R.},
        title = "{Gravitational lensing effects of vacuum strings - Exact solutions}",
      journal = {\apj},
         year = 1985,
        month = jan,
       volume = {288},
        pages = {422-427},
          doi = {10.1086/162808},
       adsurl = {https://ui.adsabs.harvard.edu/abs/1985ApJ...288..422G}
}

@Book{Chadan:1977pq,
     author    = "Chadan, K. and Sabatier, P. C.",
     title     = "Inverse Problems in Quantum Scattering Theory",
     publisher = "Springer, New York",
     year      = "1977"
}

@Book{Newton:1982qc,
     author    = "Newton, R. G.",
     title     = "Scattering Theory of Waves and Particles",
     publisher = "Springer, New York",
     year      = "1982"
}

@article{PhysRevLett.70.1739,
  title = {Stress-energy tensor of quantized scalar fields in static black hole spacetimes},
  author = {Anderson, Paul R. and Hiscock, William A. and Samuel, David A.},
  journal = {Phys. Rev. Lett.},
  volume = {70},
  issue = {12},
  pages = {1739--1742},
  numpages = {0},
  year = {1993},
  month = {Mar},
  publisher = {American Physical Society},
  doi = {10.1103/PhysRevLett.70.1739},
  //url = {https://link.aps.org/doi/10.1103/PhysRevLett.70.1739}
}

@article{PhysRevD.51.4337,
  title = {Stress-energy tensor of quantized scalar fields in static spherically symmetric spacetimes},
  author = {Anderson, Paul R. and Hiscock, William A. and Samuel, David A.},
  journal = {Phys. Rev. D},
  volume = {51},
  issue = {8},
  pages = {4337--4358},
  numpages = {0},
  year = {1995},
  month = {Apr},
  publisher = {American Physical Society},
  doi = {10.1103/PhysRevD.51.4337},
  //url = {https://link.aps.org/doi/10.1103/PhysRevD.51.4337}
}

@Article{Bordag:1994jz,
     author    = "Bordag, M.",
     title     = "Vacuum energy in smooth background fields",
     journal   = "J. Phys.",
     volume    = "A28",
     year      = "1995",
     pages     = "755",
     SLACcitation  = "%%CITATION = JPAGB,A28,755;%%"
}

@book{KL,
	Address = {Yverdon, Switzerland},
	Author = {S. G. Samko and  A. A. Kilbas and O. I. Marichev},
	Publisher = {Gordon and Breach Science},
	Title = {Fractional Integrals and Derivatives},
	Year = {1993}}

@article{Graham:2022rqk,
    author = "Graham, N. and Weigel, H.",
    title = "{Quantum corrections to soliton energies}",
    //eprint = "2201.12131",
    archivePrefix = "arXiv",
    primaryClass = "hep-th",
    doi = "10.1142/S0217751X22410044",
    journal = "Int. J. Mod. Phys. A",
    volume = "37",
    number = "19",
    pages = "2241004",
    year = "2022"
}

\end{document}